# A Micro-mechanical Modelling of the Pressure Dependence of the Void Index of a Granular Assembly:

**P. Evesque**
Lab MSSMat, UMR 8579 CNRS, Ecole Centrale Paris
92295 CHATENAY-MALABRY, France, e-mail: evesque@mssmat.ecp.fr

**Abstract:**
*This paper models the increase of density of a virgin loose granular sample submitted to a progressive compression (either isotropic or anisotropic) as an irreversible process which destroys the larger voids; a statistical mechanics approach similar to the one proposed by Boutreux & de Gennes is performed which leads to the equation of the density of normally consolidated states as a function of pressure; this equation is in agreement with experimental data and the typical variation of* e (void index) or v (specific volume) *vs.* ln(p) .

---

In a previous paper [1], it has been demonstrated using an approach à la Boltzmann that the maximum entropy principle leads to predict an exponential law for the statistical distribution of contact forces in granular materials. The exponential tail is generated by the two coupled following facts that i) it exists a discrete set of contacts with contact forces and ii) that a well defined stress is applied to the granular assembly. This stress field acts as a constraint and defines the value of the Lagrange multiplier $\beta_F$ governing the statistical distribution. This model seems to agree with numerical modelling and experimental data [2-7]. Although the modelling of [1] is concerned only with an isotropic homogeneous sample under an isotropic stress, extension to more general cases (non isotropic stress,…) is possible , for which one expects also that the force distribution decays according to such similar exponential laws with a set of different values of the Lagrange multipliers $\beta_{Fxx}$, , each one controlling the mean stress in each principal direction.

The same method was applied in [1] to find the statistics of void defects in a granular assembly; paper [1] found that the statistical distribution of void defects in granular materials shall be also exponential because the total density of the material is fixed and because the density of the medium is well defined. Although there is very little number of experimental data on the distribution of voids [8], they seem to confirm this void model. On the other hand, this void distribution (or a similar one) has been used already by Boutreux and de Gennes [9] in their approach of modelling the mechanism of compaction by the tap-tap machine (this machine is used in pharmaceutical industries to compact a sand sample with a free surface by cyclic hits of the container). In this paper, we propose to apply a similar approach to model the compaction of a loose granular material or clay submitted to a continuous increase of pressure, as a function of the pressure p. The law this model predicts is compatible with the experimental variation found for normally consolidated clays or granular materials, i.e. $e-e_o = v-v_o = -\lambda \ln(p)$, with e (v) being the void index (specific volume) and $\lambda$ being a parameter, which we try to relate to microscopic parameters. This allows to show that the $e_{min}$ introduced in [1] is probably much smaller than what can be found from experimental compaction of granular matter.

The paper is built as follows. Basic experimental behaviours obtained from soil mechanics tests [10-12] of the compaction of granular matter under isotropic and non isotropic continuous compression are recalled in the Appendix, together with some definition (critical state, void index, density, porosity). In the first part, the theory of Boutreux and de Gennes is modified and applied to the compaction of a granular medium submitted to an increase of the isotropic pressure, then to an increase of an anisotropic stress. Second part is devoted to discussion: extension to compaction of bimodal distribution is done with the help of Boutreux & de Gennes results; results are discussed in the light of typical experimental behaviours and extrapolate to broader distributions. Possible generalisation of these behaviours to clay is also discussed; distinction between natural clays, which are made of rather hard grains, and clays sedimented from solution at small concentration, whose structure is fractal [13-14], is also made. Then, links with other entropic models are drawn [15-16].

**1. Packing formed with a single kind of grains under isotropic-pressure increase**

We use now the approach proposed in [9] to describe this compaction process. The idea of [9] consists in (i) assuming that the distribution of voids of volume $\varpi$ obeys a given statistic law $\rho(\varpi)$, characterised by a parameter $\beta$ (typically $\rho(\varpi)= \beta \exp-[\beta\varpi]$), (ii) in assuming that irreversible compaction can be described by the evolution of the parameter $\beta$ only, (iii) in assuming that irreversibility is generated by the filling of large holes by some grain, so that (iv) $\beta$ varies linearly with the probability of finding holes larger than the grain size and with the increment of load ($\delta p$ here).

Let us first start with the void distribution $\rho(\varpi)$ proposed in [1] . So, following [1], the voids contained in the sample are made of an indestructible part $v_{voids-min}=e_{min} v_{solid}$ which corresponds to the minimum void-volume of the densest samples and a part which corresponds to a distribution $\rho(\varpi)$ of defects of different volume size $\varpi$. According to classical argumentation of statistical physics, one expects that this distribution $\rho(\varpi)$ shall obey a principle of maximum entropy (maximum disorder) , but it shall also respect the constraint that the sample volume v is given; in this case, the distribution $\rho(\varpi)$ is found to follow an exponential (Eq. (11 & 12 of ref. [1]):

$$\rho(\varpi) = N_\varpi/N = A\exp\{-\beta \varpi\} \qquad (1)$$

where $N_\varpi$ is the number of defects in the volume $v=(1+e)v_{solid}$ and N the total number of defects; and where A and $\beta$ are fixed by the two conditions concerning the normalisation and the mean volume: $N=\Sigma N_\varpi$ and $(e-e_{min})v_{sol}=\Sigma \varpi N_\varpi$ . So one gets $A=\beta=1/[(e-e_{min})\varpi_o]$ so that:

$$\rho(\varpi) = \beta \exp\{-\beta\varpi\} \qquad \text{with} \qquad 1/\beta = (e-e_{min})\varpi_o \qquad (2)$$

$\varpi_o$ is the typical volume of holes; from the dilatancy mechanism, it is known that the typical size of voids $\varpi_o$ is smaller than the typical grain size $\varpi_g$; however, it is not that much smaller ; so, we may expect $\varpi_o \sim\varpi_g /10$. Discussion about this value is pursued in section 2c.

Evidently, an increase of pressure perturbs the assembly, resulting in a change of the real local distribution of void defects. However, some small part only of the structure modifications corresponds to an irreversible change of the statistical distribution $\rho(\varpi)$. Let us consider nevertheless that the equation governing the new statistics remains of the type of Eq. (2) since it shall obey the principle of maximum disorder with a constraint upon volume. Let us first consider the case of a packing submitted to an isotropic stress and to an increase of this isotropic stress p. We will then consider the case of an anisotropic stress.

***1.a case of an isotropic stress.***

In order to evaluate the irreversible change of density, we proceed as recalled above and follow the procedure proposed by Boutreux & de Gennes [9]. So, compaction occurs due to the partial filling of holes larger than the grain size; the probability of occurrence of such an event is proportional to the number of holes larger than the grain size, i.e. to $\exp\{-\varpi_g/[(e-e_{min})\varpi_o]\}$, and to the pressure change $\delta p$. So

$$-\delta e = \beta^{-2}\ \delta\beta/\varpi_o = \alpha\ \delta p \int_{\varpi>\varpi_g}\rho(\varpi)d\varpi = \alpha\ \delta p\ \exp\{-\beta\varpi_g\} \qquad (3)$$

where $\alpha$ is a parameter. As $\varpi_g>>\varpi_o$, one can remark that even if $\beta$ does not vary so much this might be not the case of the exponential term which varies much faster. So, writing $e=e_o+\Delta e$, Eq. (3) leads to the approximate form:

$$-d\Delta e = \alpha\ dp\ \exp\{-\beta_o\varpi_g\}\exp\{\Delta e\ \beta_o{}^2\varpi_g\varpi_o]\} \qquad (4)$$

Let us assume that a possible solution is $\Delta e = -\lambda\ \ln(p/p_o)$ , as it is observed experimentally; this leads to $p= p_o \exp\{-\Delta e/\lambda\}$; in this case, the increment $d\Delta e$ is :

$$d\Delta e = -\lambda\ dp/p = -(\lambda /p_o)\ dp\ \exp\{\Delta e/\lambda\} \qquad (5)$$

Confrontation of Eqs. (4) and (5) leads to the identification:

$$\lambda /p_o = \alpha \exp\{-\beta_o\varpi_g\} \qquad (6a)$$

and

$$\lambda= 1/\{\beta_o{}^2\varpi_g\varpi_o\}= (e_o-e_{min})^2\varpi_o/\varpi_g \qquad (6b)$$

since

$$\beta_o{}^{-1}=(e_o-e_{min})\varpi_o \qquad (6c)$$

Eqs. (6) imposes that $p_o$ is not free but is related to $e_o$ by the relation:

$$\alpha\ p_o\ \beta_o{}^2\varpi_g\varpi_o = \exp\{\beta_o\varpi_g\} \qquad (6d)$$

$$p_o = [(e_o-e_{min})^2\varpi_o /(\varpi_g\ \alpha)]\ \exp\{\varpi_g/[(e_o-e_{min})\varpi_o]\} \qquad (6d)$$

So, in conclusion, one finds the ln(p) law found experimentally if the variation of e is small enough and if the void index remains large compared to $e_{min}$ , otherwise $\beta$ could become quite large. This implies in counter part that the void index of the uncompressible packing of grains $e_{min}$ is quite small and cannot be achieved according to usual ways.

From experimental results (see Appendix), one gets that the slope $\lambda$ is 0.3/7 about, since the variation of e is 0.3 about for a pressure change $p/p_1=10^3$. It is difficult to evaluate $e_{min}$ ; nevertheless, the densest regular packing corresponds to the porosity $\Phi=0.32$ and to $e_{min}=0.47$. This leads to evaluate $e-e_{min}=0.5$ about from the experimental data and to $\varpi_g/\varpi_o=(e_o-e_{min})^2/\lambda$ ; so, this leads to :

$$\varpi_g/\varpi_o = 6$$

which sounds reasonable.

***1.b case of an anisotropic loading at constant ratio*** $\eta=(\sigma_1-\sigma_2)/\sigma_2$

We consider the case of an anisotropic stress with axial symmetry ($\sigma_1$, $\sigma_2=\sigma_3$, $\sigma_{12}=\sigma_{23}=\sigma_{13}=0$) and a compression ($\delta\sigma_1,\delta\sigma_2,\delta\sigma_3$) which keeps constant the ratio $\eta=(\sigma_1-\sigma_2)/\sigma_2=(\sigma_1-\sigma_3)/\sigma_3=c^{ste}$. So we can parameterise the compaction with $\delta\sigma=\delta\sigma_1$ only. So considerations similar to the ones of section (1a) leads to evaluate the compaction to be governed by:

$$-\delta\Delta e = \beta^{-2}\ \delta\beta/\varpi_o \quad = \alpha'\ \delta\sigma \exp\{-\beta\varpi_g\} \qquad (7)$$

The coefficient $\alpha'$ should be approximately the same as in Eq. (3), since deformation amplitude should be similar. Again, Eq. (7) is compatible with the experimental law $\Delta e_\eta= -\lambda \ln(p_f/p_o)$ if relative variations of the void index e remain small, $p_f$ and $p_o$ being the mean final and initial stresses. In this case, one gets (see section 1a):

$$\lambda'= 1/\{\beta'_{\eta o}{}^2\varpi_g\varpi_o\}= (e'_{\eta o}-e_{min})^2\varpi_o/\varpi_g \qquad (8a)$$

$$\beta_o{}^{-1}=(e'_{\eta o}-e_{min})\varpi_o \qquad (8b)$$

$$p_o = [(e'_{\eta o}-e_{min})^2\varpi_o /(\varpi_g\ \alpha)] \exp\{\varpi_g/[( e'_{\eta o}-e_{min})\varpi_o]\} \qquad (8d)$$

This result is not in complete agreement with experimental data for the following two facts: 1) as the volume index $e'_{\eta o}$ of a granular sample submitted to an anisotropic stress (stress ratio $\eta$) is smaller than the one of a sample submitted to an isotropic stress, one should expect different values of the slope $\lambda$. This is not observed experimentally (see Appendix). 2) Another troublesome point is the fact that $\alpha$ and $\alpha'$ are determined knowing $\lambda$. In other words, the question is why the coefficients $\alpha$ and $\alpha'$ , which are coefficients which link $\delta p$ to $\delta e$, are related to the coefficient which controls the void distribution $\lambda$. Next subsection tries and proposes an alternative approach to overcome this difficulty.

***1.c alternative approach:***

This is why we may try to model the phenomena in a different way. First one remarks that there is no other reference pressure than the one applied to the sample during a compaction process; so one shall expect that the differential equation leading to compaction has no other reference state than the one at the present one. So one expects the differential equation describing the compaction to be:

$$de/e= -\lambda\, dp/p \qquad (9)$$

whose solution is:

$$\Delta e \approx -\lambda \ln(p/p_o) \qquad (10)$$

and to

$$p=p_o \exp\{-\Delta e/\lambda\} \qquad (11)$$

since the variation amplitude of the void index e remains small and since $e_o \approx 1$. So, this leads also to the behaviour observed experimentally. However, this approach does not allow to predict the value of $\lambda$, as it was done in subsections 1.a and 1.b (Eq.6b, i.e. $\lambda= [e_o-e_{min}]^2\varpi_o/\varpi_g$).

*determination of* $\lambda$

Nevertheless the theoretical value of $\lambda$ may be obtain with the approach of subsection a: let us assume for a while, that the densification process keeps the shape of the statistical distribution of voids. In this case one can assume that Eq. (2) remains valid; so, if one assumes that the irreversible evolution of the void distribution is generated by the destruction of the large holes, the way of thinking proposed by Boutreux & de Gennes [9] remains valid; this leads to Eq. (4). On the other hand, starting from Eq. (9) and replacing 1/p in Eq. (9) by the one given by Eq. (11), i.e. $\exp\{\Delta e/\lambda\}/p_o$, one gets:

$$-de= (\lambda/p_o)\, dp \exp\{\Delta e/\lambda\} \qquad (12)$$

Then, direct term-to-term identification between Eqs. (4) and (12) leads to the value of $\lambda$ given by Eqs (6b) & (8a). As $e-e_{min}$ is of the order of 0.5, and as one expects $\varpi_g/\varpi_o$ to be of order $3< \varpi_g/\varpi_o<15$ about, as mentionned in the introduction, this fixes approximately the value of $\lambda$,which is $\lambda\approx 0.06$. See section 2c for discussion.

*Remark: The case of already pre-stressed materials: the elastic behaviour*

This model applies only for a loose packing which is loaded for the first time, because it requires that noticeable irreversible compaction is generated. As a matter of fact, during a second loading or when applying stress to an already-dense material, the irreversible compaction is much smaller, and it remains smaller than the elastic behaviour, so that the model does not apply.

### *1.d increase of* $\sigma_1-\sigma_2=\sigma_1-\sigma_3$ *at constant pressure* $p=(\sigma_1+\sigma_2+\sigma_3)/3=c^{ste}$

This case of stress variation generates deformations much larger than when loading occurs at $\eta=(\sigma_1-\sigma_2)/\sigma_2$=constant, as far as $\eta$ becomes large enough.

Anyway, previous studies [17-19] show that the mechanical incremental response of the medium is isotropic at small stress . This imposes that the medium does not change of volume when p is kept constant and when the deviatoric stress ratio $\eta=(\sigma_1-\sigma_2)/p$ remains small enough (<1). However, [17-19] show that the mechanical response becomes anisotropic at larger ratio $\eta=(\sigma_1-\sigma_2)/p$. These two facts impose that the void index obeys a law of the kind:

$$e=f(1+\eta^2) \qquad (13)$$

A best fit from experimental data leads to

$$e_{\eta\neq o} - e_{\eta=0} = - \mu \ln[\{1+\eta^2/M'^2\}] \qquad (14)$$

where $\mu$ is a parameter ($\mu\approx 0.14$ ) and M' is the stress ratio $\eta$ at the critical state [10-12]. At last, one can combine Eqs. (10) & (14), and write:

$$e(\eta,p) = e_o - \lambda \ln(p/p_o) - \mu \ln[(1+\eta^2/M'^2)] \qquad (16)$$

which is a well known result for normally consolidated soil.

*Remark about the distribution of voids:*

A question arises: when $\eta$ becomes large, large deformations are generated; so one shall expect that the volume variation is not imposed by a non equilibrium process but rather by an equilibrium one: as a matter of fact, as the deformation proceeds, contacts are generated and suppressed so that holes disappear and holes are generated, so that the hole distribution shall reach an equilibrium distribution which shall reflect the micro-mechanical processes of generation and disappearance of the holes.

This can be modelled as follows. Labelling $n_\varpi$ the density of voids of size $\varpi$ ; these densities evolve according to the set of first degree equations:

$$dn_\varpi/d\varepsilon= \Sigma_{v'} \{k_{\varpi\varpi'}\, n_\varpi + k_{\varpi'\varpi}\, n_{\varpi'}\} \qquad (15)$$

Let us now consider the case of the critical state $\eta$=M'. In this case, deformation proceeds indefinitely so that $\varepsilon$ can become infinite. This leads the system of voids to reach an equilibrium without any doubt. This void distribution shall depend on the stress. And the void distribution shall correspond to a statistical equilibrium, i.e. shall obey Eq. (2).

On the other hand, experimental data show that the void ratio reached by a virgin loose packing after loading is the same whatever the stress path, i.e. $\eta=c^{ste}$ or not, as far as this one corresponds to a continuous increase of the load. This indicates that the distribution of voids shall be the same whatever the stress path. So this confirms the hypothesis made in subsections (a,b,c) which assumes that the void distribution evolves in such a way that the void distribution obeys the maximum entropy principle , i.e.Eq. (2).

## 2. Discussion and conclusion

### *2.a Packing with a distribution of grain sizes under continuous compression*

In [9], the case of the compaction of a packing made of a bimodal distribution of grains under repeated tap-tapping has been investigated. It was concluded that different cases leading to different speed of compaction shall be obtained depending on the concentration of the two different species.

For instance, if there are few large grains in a sea of small grains, the compaction is governed by the compaction of the small-grain packing; however as the isolated large grains cannot compact, the volume fraction they occupy shall not be concerned by compaction, which modifies the compaction speed.

Conversely, in the case of a large amount of large grains, the smaller grains occupy the voids in between the larger grains; in this case, the compaction process is governed by the compaction of the large-grain packing, but one shall take also into account the back-flow of small grains which are ejected when large holes are destroyed by the invasion of a large grain. This leads to predict that the slope $\lambda = \Delta e/\ln(p)$ varies with the distributions.

Let us now turn to experimental data obtained with different widths of distribution; they do show some evolution of the slope $\lambda$ with the width; however, it is rather slight for sand and other granular matter: measuring this width by the ratio $d_{60}/d_{10}$, where $d_x$ is the mesh of the sieve which sieves x% of the material, one obtains that $\lambda$ decreases slightly (from 0.11 to 0.06) when $d_{60}/d_{10}$ increases from 1 (mono disperse packing) to 10 (broad-distribution packing). Furthermore, one finds i) that the parameter $\mu$ of Eq. (16) varies also with ratio $d_{60}/d_{10}$ (it remains always around 0.1-0.2) , ii) that the initial void index $e_o$ (at a given $p_o$) decreases when increasing $d_{60}/d_{10}$ , iii), that the parameters $\lambda$ and $\mu$ does not vary anymore for $d_{60}/d_{10}$ values larger than 10. The value of $e_{min}$ decreases likely when $d_{60}/d_{10}$ increases from 1 to 10, but no precise study has been performed.

So, this modelling is at least in rather good agreement with experimental data; however theoretical modelling has still to be improved to get a better description of these well known features.

### *2.b clays*

In the same way, it is known that similar behaviours are obtained for clays contained in soils [12]; however, they are slightly different, since it is found in this case that i) the slope $\mu$ is rather independent of the material (typical value of $\mu$ is $\mu$=0.15) , ii) the slope $\lambda$ seems to depend on the maximum void index for a loose stress, the larger it is, the larger $\lambda$. For instance, $\lambda$ can be as large as 1.3 for very loose samples; in this case $e \approx 4$ at a pressure p=1bar=100kPa ; (we recall that a typical value for $\lambda$ is 0.06 for sand or dense clays). At last, correlation studies show that $\lambda$ varies linearly with the void index e at low pressure.

This last point is not in agreement with the prediction of the above model (Eq. 6b) as described in section 1, which predicts a square dependence of $\lambda$ against e , i.e. $\lambda \sim e^2$). Nevertheless one can remark that applying Eq. (6b) leads to a typical ratio $\varpi_g/\varpi_o$ ranging in the acceptable range (<15).

Furthermore, an explanation of this discrepancy can perhaps be found in the way loose clays compact: as a matter of fact, it is known that loose natural clays found in soils are made of a packing of relatively dense grains maintained in contact due to electrostatic interactions. This enables large compaction at small stress of quite loose materials. However as the stress increases, breaking of bonds between grains can occur, so that the irreversible compaction is due now to two independent different process which are competing: the first one remains the destruction of large holes, as studied in section 1, and the second one is the breaking of bonds of the skeleton of the structure. It is also possible that grain deformation plays some part in the compaction process; however this point seems to be denied by the fact that dense clays exhibit compaction behaviours similar to the one of granular materials, which seems to indicate that grains of clay are rather rigid.

For instance, experiments of pressure compaction of packing of soft grains have been also performed [12]; they do not exhibit similar behaviours. This means that the behaviours described here and in the Appendix, for which we are proposing a model, are linked to the existence of a grain-packing structure which deforms due to grain motion.

At last it is worth noting that the structure of natural clays are not identical to the one studied in [13-14] which are obtained after sedimentation of loose solutions. In this last case in particular, the aggregates have a fractal structure [13-14], which leads to a tenuous structure which breaks when increasing pressure. In this last case, the compaction during the pressure increase is linked to the shrinking and crushing of the grains and obeys a power law . This law is not compatible with the data (and the behaviours) observed in soil mechanics, and recalled in the Appendix.

### *2.c discussion about the ratio* $\varpi_g/\varpi_o$

In section 1, it has been argued that $\varpi_g/\varpi_o$ ratio shall lay between 3 & 15, since the typical volume of voids shall be much smaller than the typical grain size. However, much care has to taken with such an assumption and with its interpretation: as a matter of fact, due to the definition of void indexes e a(nd $e_{min}$), the mean void volume (and mean accessible void volume) per grain is $e\varpi_g$ (and $[e-e_{min}]\varpi_g$). So $\varpi_o$ does not represents the typical size of voids per grain for a given sample but the real typical size of the voids which enters the statistics. In this statistics the voids are assumed not to interact. An other possible interpretation is to relate the ratio $\varpi_g/\varpi_o$ to the ratio between the interaction volume required for destroying a large hole and the grain size.

### *2.d link with other statistical approach*

It is worth noting that previous models have been trying to propose an entropic approach for voids and assume a statistical distribution of the voids and of the grains in granular media; among them, one shall mention the approach of Edwards and co-workers [16], who introduced the concept of compactivity, and the one of de Larrard [17]; both of these models attempt to take into account the interaction between species explicitly [15.a, 16.a]; however, to the best of my knowledge, none of them was trying to links these evolution to a pressure variation.

On the contrary, the model proposed above is based on a statistical mechanics approach of the void distribution; it relates the evolution of this distribution to the variation of stress. It is only valid during virgin compression of a loose packing. In the case when the material is dense and/or if it was compressed previously, the compression is much less efficient leading to an irreversible compression much smaller (cf. Appendix); in this case the compression is dominated by a nearly reversible behaviour and the amplitude of deformation remains quite small.

## Appendix

Fig. A1 sketches the typical variation of the void index e of a loose granular material such as sand when increasing the mean pressure p. The dashed curve corresponds to the critical state behaviour and is obtained under a constant ratio q/p=M'. The continuous line is obtained under isotropic stress.

So, it is an experimental evidence that an increase of the mean pressure p results in an increase of the density of normally consolidated states [10-12], i.e. its mean specific volume $v=(1+e)v_{solid}$ decreases when increasing p as ln(p) in the pressure range 10kPa<p<10MPa; e is called the void index e; $v_{solid}$ is the volume occupied by the solid alone. In this formula $v_{solid}$ corresponds to a unit mass, so $v_{solid}=1$. e is related to the porosity $\Phi$ of the material by $e=\Phi/(1-\Phi)$. One considers in general that this linear behaviour $e=e_o-\lambda \ln(p)$ is the signature of the compaction of disordered packing of rather-rigid grains under pressure.

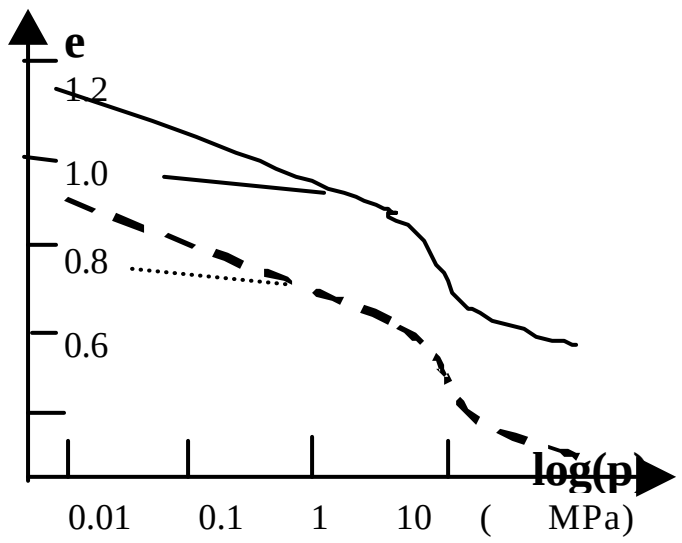

**Figure A1:** Typical experimental variation of void index $e=\Phi/(1-\Phi)$ for sand as a function of the mean pressure p in log(p) scale; — — — q/p=M' (critical state); ———— isotropic loading; $\Phi$ is the porosity. The sudden decrease is due to grain crushing.Typical values are: for 10Mpa $0.85<e_{isotropic\ loading}<1.18$ for 10kPa; for 10Mpa $0.53<e_{critcal\ state}<0.83$ for 10kPa. Typical porosity are $\Phi$=0.54 (e=1.18), $\Phi$=0.46 (e=0.85) and $\Phi$=0.33 (e=0.5)

For stress larger than 10Mpa one observes that the compaction process goes much faster; this is due to grain crushing. However, after a while a new stable statistical distribution of grain size is reached, leading to a new linear dependence of e vs. ln(p); intriguingly it exhibits the same slope (same $\lambda$) as before the crushing.

Also, experimental determination of the behaviours of samples of soft deformable elastic grains or plastic grains have been performed and typical examples can be found in [12] (pp. 202-204). They do not exhibit similar behaviours.

On the contrary, results similar to those of Fig. A1 have been obtained on natural clays. This “demonstrates” the granular nature of clay media making soils.

From Fig. A1, one can conclude that isotropic consolidation of a loose sample leads to a less dense sample than consolidation in the critical state [11-12]. This is perhaps due to the fact that the critical state is obtained after submitting the material to large deformations so that it generates much larger reorganisation of the topology of the packing contacts. However, this does not seem in agreement with the model proposed in this paper, since it assumes that void distribution is always at equilibrium.

We remark from Fig. A1 that both consolidations evolve in parallel; i.e. the slopes $\lambda$ of the lines e vs. ln(p) are the same. This means probably that the densest packing which is obtained by this method is not the densest possible one; this last one is probably much denser since it shall be quite hard to deform and it is probably not attainable by this method due to grain crushing. Finally, Fig. A1 demonstrates that crushing allows efficient compaction.

Considering now the case of an initially dense packing, or the one of a second loading of an initially loose packing, one finds that the deformation amplitude is much less and corresponds to nearly "elastic" behaviour , i.e. in the present context "elastic" means nearly reversible. This leads to nearly horizontal responses in the e vs. ln(p) plane, two examples of which are sketched in Fig. A1, starting from the normally consolidated curve and from the critical state.

*Acknoledgments:* CNES is thanked for partial funding.

The electronic arXiv.org version of this paper has been settled during a stay at the Kavli Institute of Theoretical Physics of the University of California at Santa Barbara (KITP-UCSB), in june 2005, supported in part by the National Science Fundation under Grant n° PHY99-07949.

*Poudres & Grains* can be found at :
http://www.mssmat.ecp.fr/rubrique.php3?id_rubrique=402